\documentclass[12 pt]{article}
\def\baselinestretch{1.0}
\catcode`@=11
\@addtoreset{equation}{section}
\begin{document}
\begin{center}
{ \def\baselinestretch{2.0}
\Large \bf  CHAOS AND CONTROL IN NONLINEAR BLOCH SYSTEM}
\end{center}

\vskip 20pt
\begin{center}
{\it {   Biswambhar Rakshit\footnote {e-mail: biswa\_ju2004@hotmail.com}, Papri Saha\footnote {e-mail: papri\_saha@yahoo.com} and A. Roy Chowdhury$^*$\footnote{$^*$(corresponding author) e-mail: arcphy@cal2.vsnl.net.in }  }\\
                           High Energy Physics Division \\
                             Department Of Physics     \\
                             Jadavpur University        \\
                              Kolkata - 700032\\
                                    India \\}
\end{center}
\vskip 30pt
\begin{center}
\underline{\large Abstract}\\
\end{center}
The dynamics of  two nonlinear Bloch systems is studied from the viewpoint of bifurcation and a particular parameter space has been explored for the stability analysis based on stability criterion. This enables the choice of the desired unstable periodic orbit from the numerous unstable ones present within the attractor through the process of closed return pairs. A generalized active control method have been discussed for two Bloch systems arising from different initial conditions.
	
\newpage
\section{Introduction}
The simple Bloch equations exhibit the dynamics of an ensemble of spins which do not exhibit mutual coupling. The magnetization dynamics can be determined from the linear nature of these equations. However, in high field Nuclear Magnetic Resonance(NMR), these simple structures break down due to the presence of radiation damping[1] or a demagnetizing field[2]. Manifestations of nonlinear spin dynamics could be observed due to the presence of an additional field which is proportional to the components of   the magnetization. The dipolar magnetizing field could be shown to give rise to multiple echoes in liquid helium[2] and in water at high magnetic field[3]. Moreover, radiation damping  or demagnetizing field effects [4--6] can result  in the existence of pseudo-multiquanta peaks. Complex nonlinear behaviour of the nuclear spins at high magentic field was recently demonstrated through manipulation of a large magnetization by radiation damping based electronic feedback  and giving rise to a train of steady state maser pulses of the water magnetization[7].\par Recently, Abergel[8] investigated the possibility of observing chaotic solutions of the Bloch equations. The various anomalies that arises in NMR  experiments have been studied in terms of chaos theory. U\c{c}ar et al extended the calculation of Abergel and demonstrated the synchronization of two such Bloch systems through `Master' and `Slave' arrangement through a suitably designed controller.\par
The work of OGY and Pecora and Carroll[9] led to wide applications outside the traditional scope of chaos and nonlinear dynamics. This has led to the establishment of two active areas of research viz. synchronization and control. Recent problems casts the problem of chaos synchronization  in the framework of nonlinear control theory. This unifies the study of chaos control and chaos synchronization.\par
Pyragas have proposed two methods of permanent chaos control with small time continuous perturbation in the form of linear feedback[10]. The stabilization of a unstable periodic orbit (UPO) of a chaotic system is achieved either by combined linear feedback with the use of a specially designed  external oscillator or by delayed self controlling linear feedback.\par
An open-plus-closed-loop (OPCL) method of controlling nonlinear dynamic systems was presented by Jackson and Grosu[11]. The input signal of their method is the sum of H\"{u}bler's open-loop control and a particular form of a linear closed-loop control, the goal of which can be selected as one of the UPO's embedded in the chaotic attractor or another possible smooth functions of `$t$'. The asymptotic stability of the controlled nonlinear system is realized by the linear approximation around the stabilized orbit. But the calculation of the closed-loop control signal is very difficult in some cases, specially for complex and high dimensional chaotic systems.\par
Yu. et al [12] proposed a method for controlling chaos in the form of special nonlinear feedback. The validity of this method based on the stability criterion of linear system and can be called stability criterion method(SC method). The construction of a nonlinear form of limit continuous perturbation feedback by a suitable separation of the system in the SC method does not change the form of the desired UPO. The closed return pair technique[13] is utilised to estimate the desired periodic orbit chosen from numerous UPO's embedded within a chaotic attractor.                                                                                                       The advantage of this method is that the effect of the control can be generalized directly without calculation of the maximal lyapunov exponent of the UPO using the linearization of the system. This method does not require linearization of the system around the stability orbit and  calculation of the deviation at UPO's.
The method has been used by researchers in  the control of R\"{o}ssler system, chaotic altitude motion of a spacecraft and the control of two coupled Duffing oscillators.\par
In this communication, the bifurcation analysis of the nonlinear Bloch equations in a particular parameter regime has been discussed and  stability criterion method has been used to investigate the synchronization of two such Bloch systems. The technique of generalized active control is also used for deriving a controlled trajectory of the system.
\section{Formulation}
The model is derived from a magnetization $\bf M$ precessing in the magentic induction field $\bf B_0$ in the presence of a constant radiofrequency field $\bf B_1$ with intensity ${\bf B_1}=\frac{\omega _1}{\gamma }$ and frequency $\omega _{rf}$. A magnetization-dependent  field ${\bf B}_{FB}$[14]:
$$ {\bf B}_{FB}= \gamma G{\bf M}_t e^{-i\psi },$$
with
$${\bf M}_t=M_x + iM_y$$
where $G$ is the enhancement factor with respect to the magnitude of the magnitude of the transverse magnetization and $\psi $ is the phase of the feedback field.The following modified  nonlinear Bloch equation govern the evolution of the magnetization,
\begin{equation}
\dot{M_x}=\delta M_y+G M_z(M_x \sin(\psi )- M_y\cos(\psi ))-\frac{M_x}{T _2}
\end{equation}
\begin{equation}
\dot{M_y}=-\delta M_x- \omega_1 M_z+G M_z(M_x\cos(\psi )+M_y\sin(\psi ))-\frac{M_y}{T _2}
\end{equation}
\begin{equation}
\dot{M_z}=\omega _1 M_y-G  \sin(\psi )(M_x^2+M_y^2)-\frac{(M_z-M_0)}{T_1 }
\end{equation}
where $\delta =\omega_{rf}- \omega_0$ and $T _1$, $T _2$ are the longitudinal time and transverse relaxation time respectively.The above three equation is transformed by introducing the reduced dimensionless variables: 
$$t\rightarrow  \omega _1t,\;\;\;G\rightarrow  M_0\frac{G}{\omega _1}=\gamma ,\;\;\;\delta \rightarrow \delta /\omega _1, \;\;\;T_{1,2}\rightarrow  \omega _1T_{1,2} \;\;\mbox{and}$$$$ \;\; M_x\rightarrow M_x/M_0=x\;\; M_y\rightarrow M_y/M_0=y\;\; M_z\rightarrow M_z/M_0=z.$$
The nonlinear Bloch system is then expressed as,
\begin{equation}
\dot{x}=\delta y+\gamma z(x \sin(c)-y\cos(c))-\frac{x}{\Gamma _2}
\end{equation}
\begin{equation}
\dot{y}=-\delta x-z+\gamma z(x\cos(c)+y\sin(c))-\frac{y}{\Gamma _2}
\end{equation}
\begin{equation}
\dot{z}=y-\gamma  \sin(c)(x^2+y^2)-\frac{(z-1)}{\Gamma_1 }
\end{equation}
\subsection{Analysis of the chaotic Dynamics}
The presence of the attractor has already been shown in the previous analysis of the system. The system possess two attractors for two different set of parameters viz.,  $$\gamma =35.0, \delta =-1.26, c=0.173 ,\Gamma  _1=5.0, \Gamma _2=2.5$$ and $$\gamma =10.0, \delta =1.26, c=0.7764, \Gamma _1=0.5, \Gamma _2=0.25 $$ The behaviour of the system with a change in the parameter $\gamma $ is shown with the help of bifurcation diagram. We have shown the variation of the parameter from $\gamma= 23.0 $ to $\gamma =32.0$ . The dynamics of the system is shown in Figure(1) where the system exhibits multiperiodicity for  higher values of the parameter $\gamma $ whereas it is clear that the chaoticity arises from $\gamma = 30.0$ and extends upto $\gamma =37.0$. In the range of $32.5<\gamma <33.5$, there is a sudden transition to periodicity. The unstable periodic orbit finally attains chaos which continues upto $\gamma= 37.0 $ and after that again the system intermittently transits to multiperiodic regime.
\par A detailed stability analysis based on largest lyapunov exponent of the system was carried out over a certain parameter region. The two parameters $\gamma $ and $c$ were taken into consideration and the system dynamics was distinguished into three different categories according to value of the largest positive lyapunov exponent.\par
In the stability plot the range of the x-axis and the y-axis are  $22 \leq \gamma  \leq 32 $and $-0.2 \leq c \leq 0.7$ respectively. The regions representing the dynamics of the system in the $(\gamma,c )$ plane is shown in Figure(2). The regions marked by cross `${\times}$' depicts chaotic state where largest positive lyapunov exponent $\lambda > 0.05$. A dark dot `$\bullet$' indicates  $0.003\leq \lambda \leq 0.05$. These regions exhibit multiperiodic as well as chaotic behaviour. Finally, the rest of the plane which is covered by plus sign `${\bf +}$'corresponds to purely stable behaviour with $ \lambda< 0.003$. The system illustrates a periodic behaviour or equilibrium state under these parameter conditions. 
\subsection{Control based on stability criterion}
To control the system of nonlinear Bloch equation a time continuous nonlinear dynamic system with input perturbation described by
\begin{equation}
\frac{dx}{dt}=f(x(t))+u(t)
\end{equation}
where $x \in {\bf R^n}$ and $u \in {\bf R^n}$ are the state vector and input perturbation of the system respectively. Equation (2.7) without input signal $(u=0)$ has a chaotic attractor $\Omega $. A mapping $f:{\bf R^n} \rightarrow {\bf R^n}$ is defined in n-dimensional space. We suitably decompose the function $f(x(t))$ as,
\begin{equation}
 f(x(t))=g(x(t))+h(x(t))
\end{equation}
where the function $g(x(t))=A x(t)$ is suitably disposed  as a linear part of $f(x(t),t)$ and it is required that is a full rank constant matrix, all eigenvalues of which ahve negative real parts. So the function $h(x(t))=f(x(t))-Ax$ is a nonlinear part of $f(x(t))$. Then the system  (2.7) can be rewritten as 
\begin{equation}
\frac{dx}{dt}=A x(t)+h(x(t))+u(t)\end{equation}

Let $D(x(t))=-h(x(t))$, then the function $f+D=f-h$, is a linear mapping  with respect to the state vector  x, namely,
\begin{equation}
 (f+D)(x)=A x\end{equation}

Let, $x^{\ast }(t)=x^{\ast }(t+jT),\;\; j=1,2,...$ be a period-$j$ trajectory embedded within $\Omega $. The input signal $u(t)$ is considered  as a control perturbation signal as follows
\begin{equation}
u(t)=D(x(t))-D(x^{\ast }(t))
\end{equation}
Substituting Eq.(2.11) into Eq.(2.9) system (2.7) and (2.9) can be rewritten as,
\begin{equation}
\dot{x} - \dot{x}^{\ast }=(f+D)(x)-(f+D)(x^{\ast })=A(x-x^{\ast })\end{equation}

The difference between $x(t)$ and $x^{\ast }(t)$ is defined  as an error $w(t)=x(t)-x^{\ast }(t)$, then evolution of which is determined by Eq.(2.12) as,
\begin{equation}
\dot{w}(t)=Aw(t) \end{equation}

Obviously, the zero point of $w(t)$ is its equilibrium point. Since all eigenvalues of the matrix  $A$ have negative  real parts, according  to the stability criterion of linear system, the zero point of the error $w(t)$ is asymptotically stable and $w(t)$ tends to zero when $t \rightarrow \infty $. Then the state vector $x(t)$ tends to the period-$j$ trajectory $x^{\ast }(t)$. It implies that the unstable periodic orbit is stabilized. Stable soltions belonging to different basins of initial conditions can also be the alternative solutions of very complicated periodically driven dynamic systems along with the stabilized UPO. The stabilization is obtained by modifying Eq.(2.11) as follows
$$u(t)=D(x(t))-D(x^{\ast }(t)) \hskip 150pt$$
$$=A(x-x^{\ast })+f(x^{\ast })-f(x), \;\;\;\mbox{if}\;\;\;\mid x-x^{\ast }\mid <\epsilon $$
$$=0, \hskip 150pt \mbox{otherwise}\hskip 20pt$$
For our system of equations given by (2.4--2.6) we can write the linear and nonlinear part as,
\begin{equation}
f(x)=\left(\begin{array}{ccc} -\frac{1}{\Gamma _2} & \delta &0\\ \\-\delta &\frac{1}{\Gamma _2}& -1\\ \\ 0& 1& -\frac{1}{\Gamma _1}\end{array}\right)\left(\begin{array}{c}x\\ \\y\\ \\z\end{array}\right)+\left(\begin{array}{c} \gamma z (x \sin(c)-y \cos(c))\\ \\\gamma z(x \cos(c)+y \sin(c))\\\\-\gamma \sin(c)(x^2+y^2)+\frac{1}{\Gamma _1} \end{array}\right)
\end{equation}
where
$$A=\left(\begin{array}{ccc} -\frac{1}{\Gamma _2} & \delta &0\\ \\-\delta &\frac{1}{\Gamma _2}& -1\\ \\ 0& 1& -\frac{1}{\Gamma _1}\end{array}\right)\;\;\;\;\;\;\;\;\;\;\;h(x)=\left(\begin{array}{c} \gamma z (x \sin(c)-y \cos(c))\\ \\\gamma z(x \cos(c)+y \sin(c))\\\\-\gamma \sin(c)(x^2+y^2)+\frac{1}{\Gamma _1} \end{array}\right)$$
Then the control perturbation signal is derived as follows
$$u(t)=\left(\begin{array}{c} -\gamma z (x \sin(c)-y \cos(c))+\gamma z^{\ast } (x^{\ast } \sin(c)-y^{\ast } \cos(c))\\ \\ -\gamma z(x \cos(c)+y \sin(c))+\gamma z^{\ast }(x^{\ast } \cos(c)+y^{\ast } \sin(c))\\\\\gamma \sin(c)(x^2+y^2)-\gamma \sin(c)(x^{{\ast }2}+y^{{\ast }2}) \end{array}\right)$$
In order to obtain the necessary information on an appropriate location of a desired periodic solution $x^\ast$, the strategy of the close return pair is taken into account. The method is to generate a time series of the chaotic trajectory by stroboscopically  sampling in every period $T$ when $u=0$.\\
The data sampling can be used to detect the close return pairs, which consists of two successive points nearing each other, and indicate the existence of a periodic orbit nearby. Let $x_{i,1}$  and $x_{i,2}$ are used to denote thr first point and its successive point of the $i$th  collected return pair, $=1,....,M$ respectively, where $M$ is the maximum number of collected return pairs. When the first close return pair has been detected, taking the first point  $x_{1,1}$ as a reference point, a number of close return pairs nearing the reference point can be detected.
$$\mid x_{i,1} -x_{1,1}\mid \leq \epsilon_1, \;\;\;\; \mid x_{i,2} -x_{1,1}\mid \leq \epsilon_2, \;\;\;i=1,2,......M$$ 
We define the mean value as, 
$$x^\ast=\frac{1}{2M} \sum_{i=1}^M [x_{i,1}+x_{i,2}]$$
where $x^\ast$ is regarded as an appropriate fixed point. This fixed point can be used to define a restriction condition $\mid x(t)-x^\ast(t) \mid <\epsilon $  within  which the control input signal $u \neq 0$\par
We have targeted both the attractors as described in the previous section. In both cases $M=3$.  In the first case, $x^\ast(t) =-0.221,\;\;y^\ast(t) = -0.021,\;\;z^\ast(t) =0.141\;\;\mbox{and}\;\; \epsilon =0.45$ and in the second case, $x^\ast(t) =0.219,\;\;y^\ast(t) =-0.316,\;\;z^\ast(t) =0.790\;\;\mbox{and} \;\;\epsilon =0.40$. Figures (3a) and (3b) respectively show the results of control. After a short transition both  the  attractor stabilizes on a  periodic trajectory. 
\subsection{Synchronization using generalized active control}
Bai and Lonngren[15] proposed an active control process and two Lorenz systems were synchronized using that technnique. Ho and Hung used the method of generalized active control to synchronize two completely different systems[16]. Here we show the synchronization of two nonlinear Bloch equations and this technique is different from the one by U\c{c}ar et al[17] in the sense that the control signals in this particular case do not contain any positive gain term. It is found to be equally effective and the results can be compared to that obtained earlier. A second nonlinear Bloch equation is considered with the same parameter values as the previous one but differing in the initial conditions.
\begin{equation}
\dot{u}=\delta v+\gamma w(u \sin(c)-v\cos(c))-\frac{u}{\Gamma _2}+\eta_1(t)
\end{equation}
\begin{equation}
\dot{v}=-\delta u-w+\gamma w(u\cos(c)+v\sin(c))-\frac{v}{\Gamma _2}+\eta _2(t)
\end{equation}
\begin{equation}
\dot{w}=v-\gamma  \sin(c)(u^2+v^2)-\frac{(w-1)}{\Gamma_1 }+\eta _3(t)
\end{equation}
Subtracting Equations(2.15--2.17) from (2.4--2.6) and performing the required calculations, the control signals are obtained as follows
\begin{equation}
\eta_1(t)=\zeta_1(t)-\gamma w(u \sin(c)-v \cos(c))+\gamma z(x\sin(c)-y \cos(c))
\end{equation}
\begin{equation}
\eta_2(t)=\zeta_2(t)-\gamma w(u \cos(c)+v \sin(c))+\gamma z(x\cos(c)+y \sin(c))
\end{equation}
\begin{equation}
\eta_3(t)=\zeta_3(t)+\gamma  \sin(c)(u^2+v^2)-\gamma \sin(c)(x^2+y^2)
\end{equation}
where
$$\zeta _1(t)= \left(\frac{1}{\Gamma _2}-1\right)\varepsilon _1- \delta \varepsilon _2$$
$$\zeta _2(t)= \delta \varepsilon _1+ \left(\frac{1}{\Gamma _2}-1\right)\varepsilon _2+\varepsilon _3$$
$$\zeta _3(t)=  \left(\frac{1}{\Gamma _1}-1\right)\varepsilon _3-\varepsilon _2$$
with $\varepsilon _1=u-x$, $\varepsilon _2=v -y$, $\varepsilon _3=w-z$. The parameters are $\gamma = 35.0, \delta =-1.26, c = 0.173 ,\Gamma _1=5.0, \Gamma _2 = 2.5$ and the error dynamics is shown in Figure (4) which converges  zero indicating the synchronization between the two systems which evolve from two different states.
\section{Conclusion}
In our above analysis  we have studied  a different methodology of synchronization in a Maxwell-Bloch system by analyzing its bifurcation pattern and stability. The main emphasis is on identifying a unstable periodic orbit and to adopt the strategy of closed return pairs to control whose effectiveness is represented in our results. Regarding the method of active control, while the previous authors  were concerned about two systems with different parameter values our consideration is on two different initial state of the system--that is their magnetization. Our approach does not take into account  any gain term as it is more generalized than the previously considered one.\\
\\
{\bf Acknowledgement:}: One of the authors B.Rakshit is thankful to CSIR, Govt. of India for a research fellowship in a project.

\section{References}
{[1]}.  N. Bloembergen and R.V. Pound - {\it Phys. Rev. E} {\bf \underline {95}}  {(1954)} {8}.\\
{[2]}.  G. Deville, M. Bernier and J.M. Delrieux - {\it Phys. Rev. B} {\bf \underline {19}}  {(1979)} {5666}.\\
{[3]}.  R. Bowtell, R.M. Bowley and P. Glover - {\it J. Mag. Reson.} {\bf \underline {88}}  {(1990)} {643}.\\
{[4]}.  Q. He et al. - {\it J. Chem. Phys.} {\bf \underline {98}}  {(1993)} {6779}.\\
{[5]}.  J. Jeener, A. Vlassenbroek and P. Broakaert - {\it J. Chem. Phys.} {\bf \underline {103}}  {(1995)} {1309}.\\
{[6]}.  J. Jeener - {\it Phys. Rev. Lett.} {\bf \underline {82}}  {(1999)} {1772}.\\
{[7]}.  D. Abergel, A. Louis-Joseph and J.Y. Lallemand - {\it J. Chem. Phys.}\\ {\bf \underline {116}}  {(2002)} {7073}.\\
{[8]}.  D. Abergel - {\it Phys. Lett. A} {\bf \underline {302}}  {(2002)} {17}.\\
{[9]}.  L.M. Pecora and T.L. Carroll - {\it Phys. Rev. Lett} {\bf \underline {64}}  {(1990)} {821}.\\
{[10]}.  K. Pyragas - {\it Phys. Lett A} {\bf \underline {170}}  {(1992)} {421}.\\
{[11]}.  E. Atlee Jackson and I. Grosu  - {\it Physica D} {\bf \underline {85}}  {(1995)} {1}.\\
{[12]}.  H.J.Yu,Y.Z.Liu and I.H.Peng - {\it Phys.Rev.E } {\bf \underline {69}}  {(2004)} {066203}.\\
{[13]}.  E.R. Hunt - {\it Phys. Rev. Lett} {\bf \underline {68}}  {(1992)} {1259}.\\
{[14]}.  R.F. Hobson and R. Kayser - {\it J. Mag. Reson.} {\bf \underline {20}}  {(1975)} {458}.\\
{[15]}.  E.W. Bai and K.E. Lonngren - {\it Chaos, Solitons \& Fractals} {\bf \underline {8}}  {(1997)} {51}.\\
{[16]}.  M.C. Ho and Y.C. Hung - {\it Phys.  Lett. A} {\bf \underline {301}}  {(2002)} {424}.\\
{[17]}.  A. U\c{c}ar, K.E. Lonngren and E.W. Bai - {\it Phys. Lett. A} {\bf \underline {314}}  {(2003)} {96}.\\
\section{Caption for Figures}
{[1]}.  The bifurcation diagram of the Bloch system with respect to the parameter $\gamma $.\\
{[2]}.  Chaotic domain in the $(\gamma , c )$ plane. `$\bullet $' and `$+$' represent stable states whereas `$\times$' denotes chaos.\\
{[3]}.  Temporal evolution of the system after the application of the control perturbation due to stability criterion. a) $\gamma =35.0, \delta =-1.26, c=0.173 ,\tau _1=5.0, \tau _2=2.5$; b)$\gamma =10.0, \delta =1.26, c=0.7764, \tau _1=0.5, \tau _2=0.25 $; .\\
{[4]}.  Error dynamics of the system due to control by generalized  active control.\\

\end{document}